\documentclass[aps,prl,amsmath,amssymb,footinbib,showpacs,twocolumn,superscriptaddress,longbibliography]{revtex4-2}
\usepackage{amsmath}
\usepackage{flushend}
\usepackage{amssymb}
\usepackage{amsthm}
\usepackage[toc,page]{appendix}
\usepackage{newtxmath}
\usepackage{setspace}
\usepackage{graphicx}
\usepackage{braket}
\usepackage{mathrsfs}
\usepackage{dcolumn}
\DeclareUnicodeCharacter{2212}{-}
\usepackage{float}
\usepackage[colorlinks = true,linkcolor = red,urlcolor  = blue,citecolor = blue,anchorcolor = blue]{hyperref}
\usepackage[utf8]{inputenc}
\usepackage[english]{babel}
\usepackage{bm}
\usepackage[caption=false]{subfig}

\begin{document}

\title{Topological invariant for finite systems in the presence of disorder}

\author{Robert Eissele}
\affiliation{Department of Physics and Astronomy, West Virginia University, Morgantown, WV 26506, USA}

\author{Binayyak B. Roy}
\affiliation{Department of Physics and Astronomy,
Clemson University, Clemson, SC 29634}

\author{Sumanta Tewari}
\affiliation{Department of Physics and Astronomy,
Clemson University, Clemson, SC 29634}

\author{Tudor D. Stanescu}
\affiliation{Department of Physics and Astronomy, West Virginia University, Morgantown, WV 26506, USA}

\begin{abstract}
Topological invariants, rigorously defined only in the thermodynamic limit, have been generalized to topological indicators applicable to finite-size disordered systems. However, in many experimentally relevant situations, such as semiconductor–superconductor (SM–SC) hybrid nanowires hosting Majorana zero modes, the interplay between strong disorder and finite-size effects renders these indicators (e.g., the so-called topological visibility) biased and ill-defined, significantly limiting their usefulness. In this paper, we propose the topological invariant rigorously defined for an infinite system constructed by periodically repeating the original finite disordered system, as a topological indicator. Using the one-dimensional SM–SC hybrid nanowire as an example, we show that this general and transparent approach yields faithful topological indicators free from the biases affecting commonly used finite-size indicators, capturing the nature (topological or trivial) of the phase at generic points in parameter space, and providing a reliable tool for interpreting experimental results.
\end{abstract}

\maketitle

\textit{Introduction:} Topological phases and their associated invariants are rigorously defined only in the thermodynamic limit, e.g., in clean infinite systems, where (crystal) momentum serves as a well-defined quantum number. Under these idealized conditions, topological invariants for superconductors/superfluids,  typically quantized as integers ($\mathbb{Z}$) or binary values ($\mathbb{Z}_2$), 
and calculated by evaluating certain functions of the crystal momentum at specific values of the momentum or integrating them over the Brillouin zone, 
predict the existence of robust boundary-localized states called Majorana zero modes (MZMs) via the bulk-boundary correspondence \cite{KitaevPU2001,ReadGreenPRB2000,FuKanePRL2008,Majorana1937,WilczekPRL1982,NayakNPB1996,NayakRMP2008}. The notion of topological invariant has been generalized to realistic conditions by formulating operational definitions of so-called {\em topological indicators} applicable in real space, particularly useful for characterizing finite disordered systems \cite{FulgaPRB2012,AkhmerovPRL2011}. However, the finite-size generalizations are problematic, as the intrinsic ambiguity underlying the operational definition of a ``topological phase'' in finite-size systems 
is reflected by the topological indicators no longer being quantized, e.g., taking continuous values between $\pm 1$. 

An example of practical importance is provided by the one-dimensional (1D) topological superconductor realized in spin-orbit coupled semiconductor-superconductor (SM-SC) heterostructures \cite{SauPRL2010,SauPRB2010,LutchynPRL2010,OregPRL2010,Tewari2010219,mourik2012signatures, Deng2012, Das2012, rokhinson2012fractional, churchill2013superconductor, finck2013anomalous, deng2016majorana, zhang2017ballistic, chen2017experimental, nichele2017scaling, albrecht2017transport, o2018hybridization, shen2018parity, 
sherman2017normal, vaitiekenas2018selective, albrecht2016exponential, PhysRevB.107.245423}, which is being investigated as a possible platform for topological quantum computation (TQC) with Majorana zero modes (MZMs) \cite{Microsoft2025}. In this case, particle-hole symmetry and the unitarity of the zero-bias reflection matrix $r$ in the thermodynamic limit (i.e., for vanishing transmission) allow the definition of a discrete topological invariant $\mathcal{Q}_0 = \det(r)$ taking values  $\mathcal{Q}_0 = +1$ (trivial phase) and $\mathcal{Q}_0 = -1$ (topological SC phase) \cite{Akhmerov_2011,Fulga_2011,Fulga_2012}.  
In realistic finite-length systems unitarity is lost and we have $-1 \leq \det(r) \leq 1$, with $\det(r) <0$ typically interpreted as indicating a topological SC phase \cite{JayDSau2016}. However, as discussed in detail in a recent study \cite{Day2025}, the scattering invariant of a finite system is plagued with various biases. In the tunneling regime, for example, the boundary Majorana modes may couple (generating finite energy splittings) and the system appears trivial, i.e., $\det(r)>0$. Calculating $\det(r)$ in the presence of a small dissipative term that mimicks the presence of additional quasiparticle sources or sinks, i.e., introducing the so-called {\em topological visibility} (TV) \cite{pikulin2021protocolidentifytopologicalsuperconducting,PhysRevB.107.245423}, can alleviate this issue in short systems, but generates other problems, in particular false positives associated with the presence of so-called partially separated Andreev bound states (ps-ABS) or quasi-Majorana modes induced by disorder \cite{moore2018two,Moore2018,vuik2018reproducing}.


In this paper, we propose a general, straightforward, and physically-transparent approach to constructing topological indicators for finite disordered systems free of the biases discussed in the literature \cite{Day2025}. We propose to use as indicator the topological invariant rigorously defined for the infinite system obtained by periodically repeating the original disordered system in a superlattice. The approach can be implemented in arbitrary spatial dimensions and the resulting invariant reflects the bulk properties of the original system, taking a specific value for the entire class of disordered systems that generate the same infinite periodic system. Considering the relevant example of one-dimensional topological superconductors based on the SM-SC heterostructure, we illustrate the construction of the ``periodic disorder'' invariant (PDI) and investigate the impact of increasing the size of the disordered system on the topological phase diagram obtained from the PDI. By explicitly calculating the low-energy spectral and real-space properties of the finite system, we demonstrate that the PDI can be used as a faithful topological indicator. In particular, we show that the invariant can unambiguously distinguish robust Majorana modes from disorder-induced trivial quasi-Majorana states, eliminating the false positive (or negative) predictions based on the topological visibility. Ultimately, we show that the PDI approach is capable of generating well-defined topological maps free of the obvious biases \cite{Day2025}, with parameter values corresponding to robust (operationally-defined) topological phases located far enough from the phase boundary, and hence should be the appropriate topological indicator used to analyze the experimental phase diagram obtained, e.g., from the differential conductance \cite{PhysRevB.107.245423,Microsoft2025}.

{\textit{Model Hamiltonian and the Winding Number PDI Invariant:}} Consider a 1D SM-SC hybrid system described by a tight-binding model defined on a lattice with lattice constant $a$ containing $L\cdot N$ sites divided into $N$ identical ``cells'' (or ``segments''), each containing $L$ sites. The thermodynamic limit corresponds to $N \!\rightarrow\!\infty$. For a clean system, after imposing periodic boundary conditions, the zero frequency Green's function has the form $G(0,k)=-[H(k)]^{-1}$, with
\begin{equation}
 H(k)=h_0+h_1e^{ik}+h^\dagger_1e^{-ik}, \label{Hk}
\end{equation}
where $k$ is the wave vector in units of $1/a$, $h_0 = (2t-\mu)~\!\sigma_0 \tau_z +\Gamma ~\!\sigma_x \tau_z -\gamma~\!\sigma_y\tau_y$ is the on-site contribution, and $h_1= -t ~\!\sigma_0 \tau_z -i \alpha/2 ~\!\sigma_y\tau_z$ designates the nearest-neighbor contribution. Here, $t$ is the nearest-neighbor hopping, $\alpha$ the Rashba spin-orbit coupling, $\mu$ the chemical potential, $\Gamma$ the (longitudinal) Zeeman field, $\gamma$ the effective SM-SC coupling, $\sigma_0$ and $\tau_0$ are $2\times2$ identity matrices, while ${\bm \sigma}=(\sigma_x, \sigma_y, \sigma_z)$ and ${\bm \tau}=(\tau_x, \tau_y, \tau_z)$ are Pauli matrices associated with the spin and particle-hole degrees of freedom, respectively. In the presence of a disorder potential $V_{dis}(i) = V_{dis}(i+(p-1)L)$, where $i=1,\dots,L$ labels the cell sites and $p=1, \dots, N$ labels different cells, the (Fourier transformed) zero frequency Green's function matrix is $\widehat{G}(0,q)=-[\widehat{H}(q)]^{-1}$, where $\widehat{H}(q)$ contains blocks of the form
\begin{eqnarray}
\widehat{H}_{ij}(q) &=& [h_0 + V_{dis}(i)~\!\sigma_0\tau_z]~\! \delta_{ij}+h_1~\! \delta_{j,i+1}+h_1^\dagger~\! \delta_{j,i-1} \nonumber \\
&+& h_1e^{i q}~\! \delta_{i,N}\delta_{j,1}+h_1^\dagger e^{-i q}~\! \delta_{i,1}\delta_{j,N}, \label{hatH}
\end{eqnarray}
with $q$ being the wave vector within the reduced Brillouin zone in units of $1/La$. For the clean system, the winding number invariant $\nu = \int_{-\pi}^{\pi}\frac{dk}{4\pi i} {\rm Tr}\left[S G^{-1}(0,k) ~\!\partial_kG(0,k)\right]$, where $S = \sigma_0\tau_x$, can be written in terms of the nearest neighbor contributions to the Hamiltonian and the Green's function as
\begin{equation}
\nu = \frac{1}{2}{\rm Tr}\left[S(h_1g_1^\dagger-h_1^\dagger g_1)\right], \label{nu}
\end{equation}
where $g_1=\int_{-\pi}^\pi \frac{dk}{2\pi} ~G(0,k) e^{-ik}$ is the nearest-neighbor Green's function. Remarkably, a similar expression holds for the disordered system, with $g_1$ being replaced by the position-averaged nearest-neighbor Green's function 
\begin{equation}
\overline{g}_1=\int_{-\pi}^\pi \frac{dq}{2\pi} \frac{1}{L}\left[\sum_{i=1}^{L-1}\widehat{G}_{i,i+1}(0,q) + e^{-iq} \widehat{G}_{L,1}(0,q)\right].  \label{barg1}
\end{equation}
Thus, Eq. (\ref{nu}), with $g_1\rightarrow \overline{g}_1$ given by Eq. (\ref{barg1}), represents the winding number PDI for a 1D disordered superconductor. Calculating $\widehat{G}(0,q)$ using an implementation of the recursive Green's function method 
enables us to efficiently map the relevant parameter space for large systems and multiple disorder realizations. In this study we calculate topological maps in the Zeeman field--chemical potential plane for a strongly coupled SM-SC system with $\gamma=2\Delta_0=0.5~$meV and SM model parameters consistent with InAs \cite{PhysRevB.107.245423,Microsoft2025}. The trivial and topological regions correspond to $\nu=0$ and $\nu=1$, respectively.

Similarly, one can construct the Kitaev invariant of the 1D superconductor \cite{KitaevPU2001}, which for a clean system is given by the sign of the product of the Pfaffians of the skew-symmetric matrices $H(k) S$ at $k=0$ and $k=\pi$. The corresponding PDI has the form
\begin{equation}
{\cal M} = {\rm sign}\left\{{\rm Pf}\left[\widehat{H}(0)\widehat{S}\right]\right\}~\! {\rm sign}\left\{{\rm Pf}\left[\widehat{H}(\pi)\widehat{S}\right]\right\},
\end{equation}
where $\widehat{H}(q)$ is given by Eq. (\ref{hatH}) and $\widehat{S}$ is a block-diagonal matrix with $S = \sigma_0\tau_x$ diagonal contributions. Note that, since $e^{\pm i q}=+1$ ($-1$) for $q=0$ ($q=\pi$), the two factors correspond to periodic and anti-periodic boundary conditions, respectively. Finally, we point out that, following arguments similar to those in Ref.~\onlinecite{Ghosh_2010}, the Kitaev PDI can be expressed in terms of the winding number PDI as ${\cal M} = (-1)^\nu$. 


\begin{figure}[t]
\centering
\includegraphics[width=0.48\textwidth]{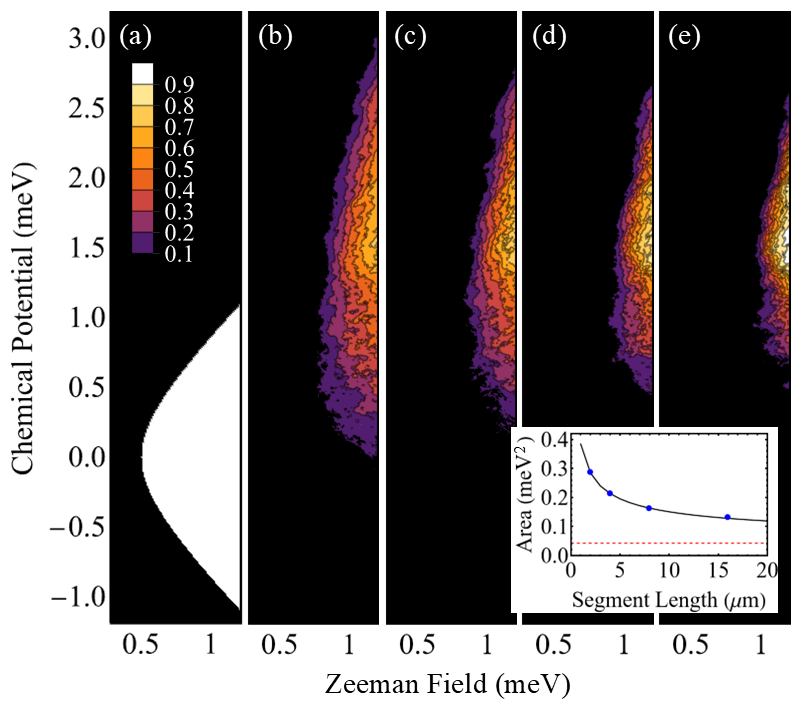}
\caption{(a) Topological phase diagram for a clean, infinite system. The black and white regions correspond to the topologically trivial and nontrivial phases, respectively. (b)--(e) Disorder-averaged topological phase diagrams for systems with increasing cell size: (b) $L=2~\mu$m, (c) $L=4~\mu$m, (d) $L=8~\mu$m, (e) $L=16~\mu$m. The contours correspond to a $10\%$ increase of the probability of having a topological superconducting phase, from nearly zero (black) to over $90\%$ (white). The averaging involves $100$ disorder realizations with a random potential amplitude $V_0=1.5~$meV. The (average) area of the topological region decreases with $L$ (see blue dots in the inset) approximately as $A_\infty + \kappa/\sqrt{L}$ (black line), with $A_\infty\approx 0.042~$meV$^2$ (dashed red line).}    \label{Fig1}
\end{figure}

\begin{figure}[t]
\centering
\includegraphics[width=0.42\textwidth]{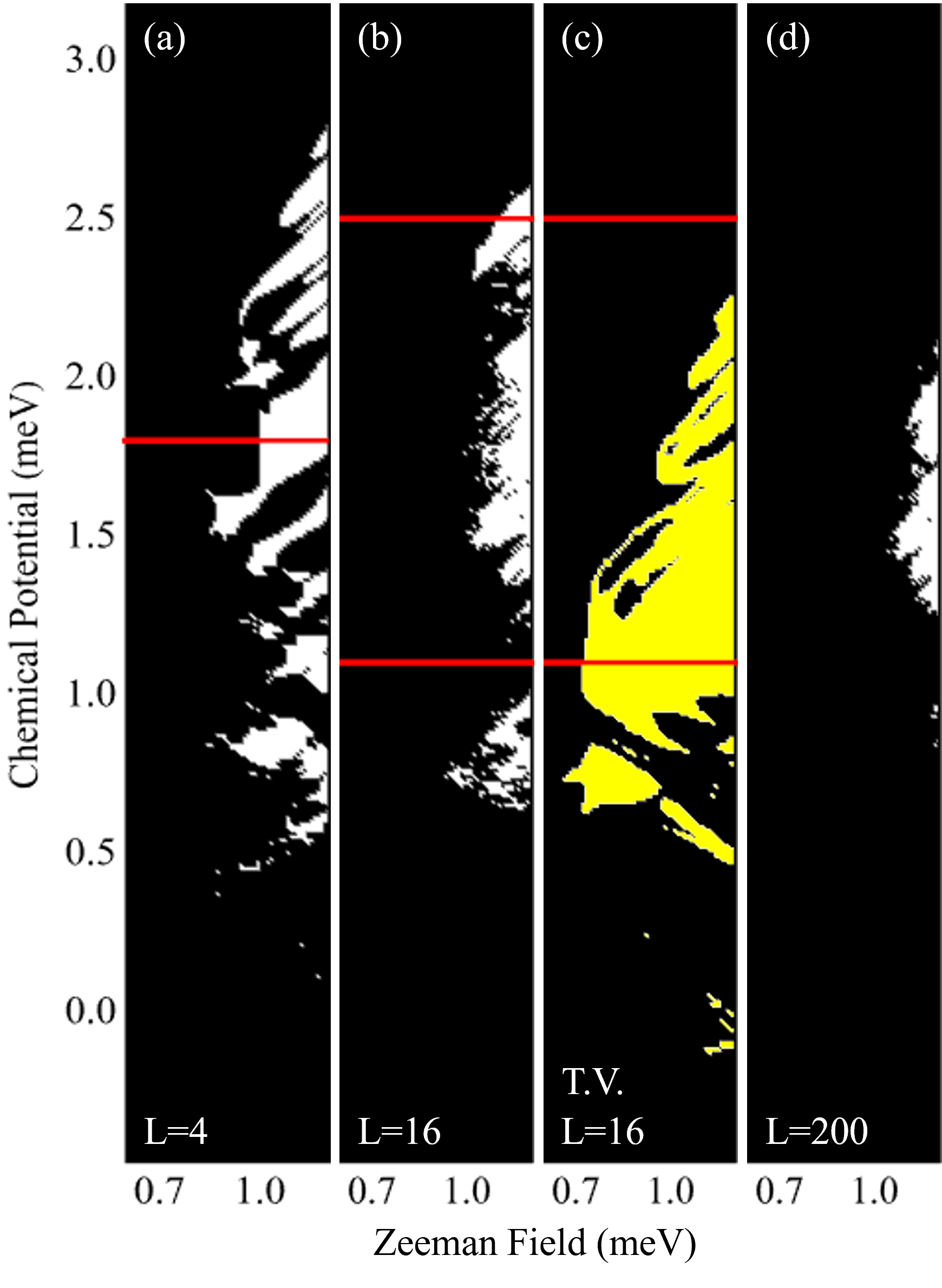}
\caption{Topological phase diagrams corresponding to specific disorder realizations and cell lengths (a) $L=4~\mu$m, (b) $L=16~\mu$m, and (d) $L=200~\mu$m. (c) Topological visibility for a wire of length $L=16~\mu$m corresponding to a single cell from the system shown in (b). The low-energy spectral features along the cuts marked by red lines are shown in Figs. \ref{Fig3} and \ref{Fig4}. Note that the correlation between the topological region in (d) and the disorder-averaged topological phase diagram in Fig. \ref{Fig1}(e) is $54\%$.}    \label{Fig2}
\vspace{2mm}
\end{figure}

\textit{Topological phase diagram from PDI:} We first investigate the dependence of the topological phase diagrams constructed using the winding number PDI on the wire length $L$. In particular, assuming that $\nu=1$ indicates an operationally topological phase, we examine the meaningfulness of the following question: If a disordered wire of length $L$ is in a topological phase ($\nu=1$), will this state be robust against increasing the system size? Since this involves ``adding'' segments having the same ``type'' of disorder, but arbitrary realizations (or configurations), we address the problem by calculating the disorder-averaged topological phase diagrams for systems of different lengths. Specifically, for a given $L$ value we consider 100 different disorder realizations corresponding to random potentials having $\langle V_{dis}\rangle =0$ and $\langle V^2_{dis}\rangle =V_0^2$, with an amplitude $V_0=1.5~$meV. 
The disorder-averaged maps for $L=2, 4, 8,~{\rm and}~ 16~\muup$m are shown in Fig. \ref{Fig1}. Two examples of contributions from specific disorder realizations are provided in Fig. \ref{Fig2}(a) and (b). For reference, in Fig. \ref{Fig1}(a) we show the topological phase diagram of the clean system.

For a given disorder configuration, the topological map [see, e.g., Fig. \ref{Fig2}(a) and (b)] consists of a trivial region ($\nu=0$, black) and a (highly fragmented) topological region ($\nu=1$, white). In general, different disorder configurations correspond to different topological maps. Consequently, for a given set of control parameters, ($\Gamma, \mu$), averaging over multiple disorder realizations results in a value $\langle \nu \rangle$ between $0$ and $1$, which can be interpreted as the probability for a wire (of given length, disorder amplitude, and control parameter values) to be in a topological phase. Thus, we can view panels (b-e) of Fig. \ref{Fig1} as showing probability maps for systems of increasing length $L$. 

The first notable result revealed by Fig. \ref{Fig1} is that, in the presence of disorder, the parameter space region characterized by a non-negligible probability of supporting a topological SC phase is ``shifted'' with respect to its clean counterpart toward larger values of $\mu$ and $\Gamma$. In particular, for $V_0=1.5~$meV most of the topological region characterizing the clean system is likely to support trivial states. The second important feature is the reduction of the area characterized by ``intermediate'' values of the probability (e.g., $0.1 \lesssim \langle \nu \rangle \lesssim 0.8$) with increasing $L$, together with a corresponding enhancement of the maximum probability, which is associated with a (relatively small) region around $\mu \approx 1.5~$meV having $\Gamma \gtrsim 1.1~$meV. This indicates that, on the one hand, the ``fragmentation'' of the topological region is associated with finite size effects (i.e., it is alleviated by increasing the wire length $L$) and, on the other hand, topological states occurring within the high probability region are more robust against increasing the wire length than the topological states occurring outside this region. For example, if a wire of length $L$  is topological for $(\Gamma, \mu) = (1.15, 1.5)~$meV, it is likely to remain topological when extended to a length $L + L^\prime$, with $L^\prime$ arbitrary. On the other hand, if a wire of length $L$  is topological for $(\Gamma, \mu) = (0.9, 2.5)~$meV, the extended system of length $L + L^\prime$ is likely to support a trivial SC phase, with a probability that approaches $1$ when $L^\prime \rightarrow \infty$. This, however, does not necessarily make the operationally defined phase of the length--$L$ system at and near $(\Gamma, \mu) = (0.9, 2.5)~$meV less ``topological'' than the phase emerging at and near $(\Gamma, \mu) = (1.15, 1.5)~$meV (see below). We note that stability/instability of the topological state against enlarging the system size is consistent with the length dependence of the average area of the topological region (see the inset of Fig. \ref{Fig1}). In particular, within the high probability region, a small topological phase survives at arbitrarily large $L$ values, as shown explicitly by the results in Fig. \ref{Fig2}(d) for a system with $L=200~\muup$m obtained by concatenating all $100$ wires used for calculating the probability map in Fig. \ref{Fig1}(b).
Finally, this analysis emphasizes the practical importance of determining the probability map for (nominally identical) systems of given size $L$: If a high probability region (with $\langle \nu \rangle$ significantly above $0.5$) is not experimentally accessible (e.g., one cannot access the regime $\Gamma > 1.1~$meV), increasing the size of the system is pointless, as it will not generate more robust topological states.     

Next, we address the critical question concerning the correlation between a system being characterized by a nontrivial PDI value and the possibility of defining (operationally) a topological phase. Here, we adopt a qualitative perspective and identify a topological phase as being associated with a finite control parameter region supporting a robust near-zero energy pair of Majorana modes localized near the ends of the system. Of course, a full operational definition would involve specific quantitative bounds for the Majorana energy splitting, separation length, overlap, etc. As an example, let us focus on the constant-$\mu$ cut corresponding to the red horizontal line in Fig. \ref{Fig2}(a).
\begin{figure}[t]
\centering
\includegraphics[width=0.42\textwidth]{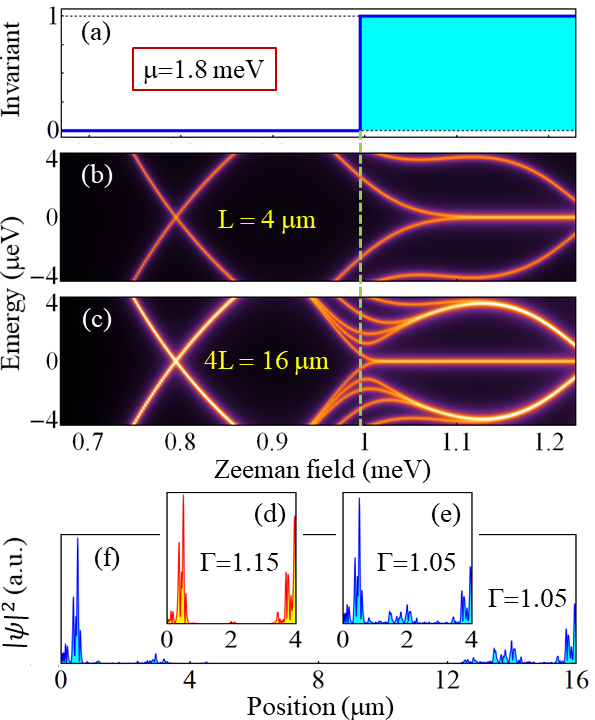}
\caption{(a) Topological invariant along the cut in Fig. \ref{Fig2}(a). (b) Low-energy DOS corresponding to a single segment of length $L=4~\mu$m. The x-like feature is generated by a (topologically trivial) ABS. The robust zero-energy feature corresponds to a pair of well separated Majorana modes localized near the ends of the wire, as shown in (d) for a Zeeman field $\Gamma=1.15~$meV. Upon approaching the phase boundary, the characteristic length scale of the Majorana modes increases [see panel (e)] and they hybridize acquiring a finite energy. (c) Low-energy DOS for system consisting of four (identical) segments. The leftmost and rightmost Majorana modes survive [see panel (f)], while ``internal'' Majorana pairs hybridize and acquire finite energy. (d)-(f) Position dependence of the lowest energy modes in (b) and (c) corresponding to a Zeeman field $\Gamma$ (given in meV).}    \label{Fig3}
\vspace{2mm}
\end{figure}
The PDI is $\nu=0$ for Zeeman field values $\Gamma \lesssim 1~$meV and $\nu =1$ at larger fields [also see Fig. \ref{Fig3}(a)]. The field dependence of the low-energy density of states (DOS) for the corresponding $L=4~\muup$m wire, shown in Fig. \ref{Fig3}(b), reveals the presence of a robust near-zero energy state at $\Gamma > 1~$meV, which splits upon approaching $\Gamma\approx 1~$meV from above. This corresponds to a pair on Majorana modes with a characteristic length scale (on the order of $1-2~\muup$m) that increases as one approaches  $\Gamma\approx 1~$meV from above [see Fig. \ref{Fig3}(d) and (e)]. By contrast, for $\Gamma < 1~$meV the system is gapped, with the exception of an almost-linearly dispersing state near $\gamma\approx 0.8~$meV, which is an Andreev bound state (ABS) localized in the middle of the wire (not shown). We conclude that the $\nu=0$ region is, indeed, topologically trivial, while a topological phase can be operationally defined within the $\nu=1$ region, sufficiently far from the ``transition'' point at $\Gamma\approx 1~$meV. We have numerically confirmed this correlation for systems having different lengths and multiple disorder realizations.  

\begin{figure}[t]
\centering
\includegraphics[width=0.45\textwidth]{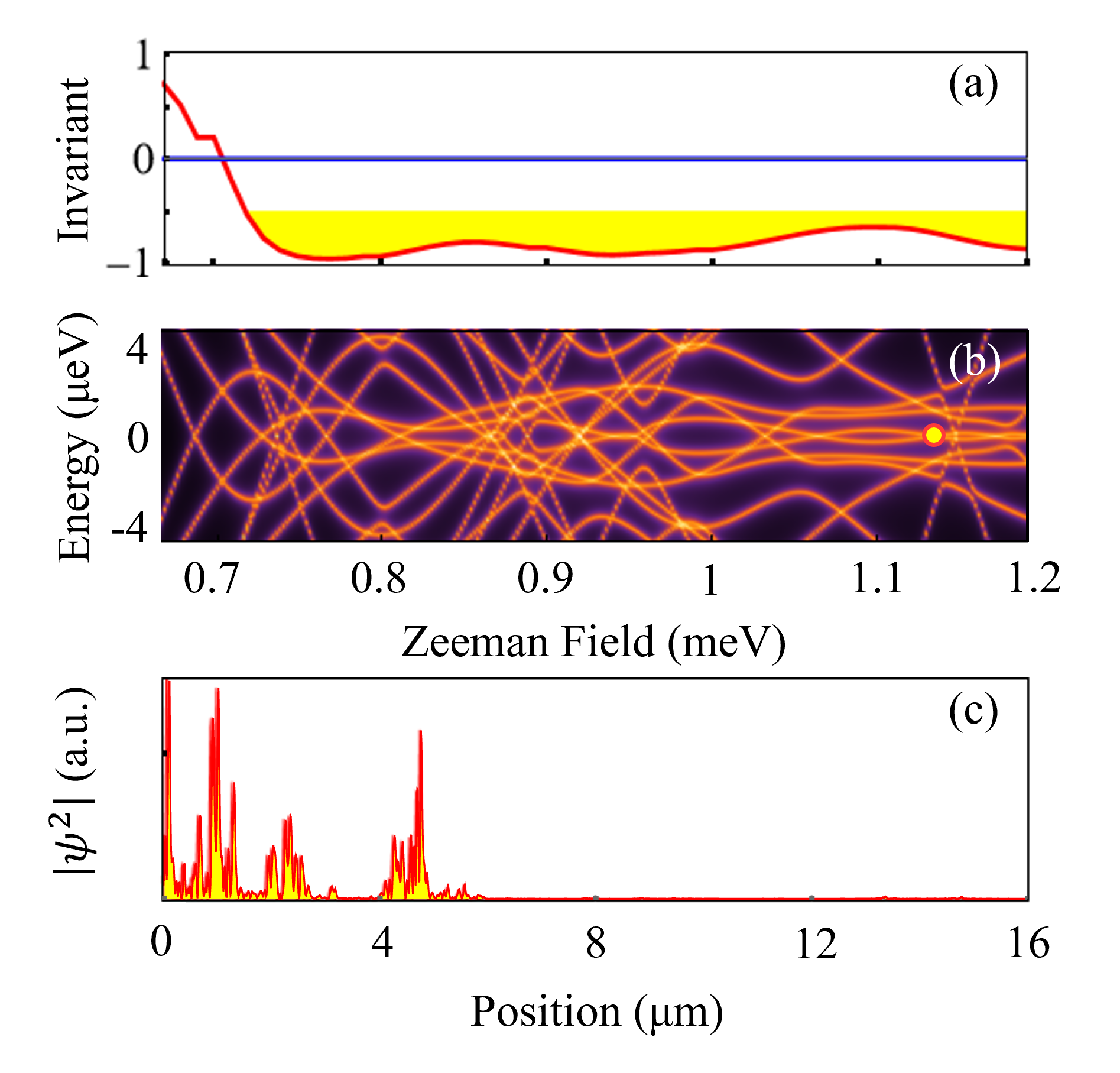}
\vspace{-4mm}
\caption{(a) Topological invariant (blue) and topological visibility (red) along the lower cuts in Fig. \ref{Fig2}(b) and (c) corresponding to $\mu=1.1~$meV. (b) Low-energy DOS corresponding to a single segment. Note the absence of a robust zero-energy mode, despite the TV predicting a topological phase for $\Gamma \gtrsim 0.73~$meV. (c) Lowest energy state for $\Gamma=1.15~$meV consisting of a pair of partially-separated Majorana modes localized near the left end of the system.}    \label{Fig4}
\end{figure}

To gain further insight into PDI construction, we consider a longer wire obtained by concatenating four $L=4~\muup$m segments. The corresponding DOS, shown in Fig. \ref{Fig3}(c), shows a robust near-zero energy mode extending through almost the entire $\nu=1$ region. In addition, one can clearly notice a set of modes that almost close the bulk gap near $\Gamma\approx 1~$meV, a feature that is absent in panel (b) due to large finite size effects. The lowest energy state at $\Gamma=1.05~$meV consists of two Majorana modes [see Fig. \ref{Fig3}(f)] with spatial profiles almost identical to those in Fig. \ref{Fig3}(e). The inter-segment Majoranas (e.g., the right Majorana of segment $1$ and the left Majorana of segment $2$, etc.) have hybridized and acquired finite energy. This provides an intuitive picture of the PDI construction: A system is in a topological phase (with $\nu=1$) if it supports pairs of near-zero energy Majorana modes with the property that the intra-segment Majorana coupling is weaker than the inter-segment coupling --- a picture directly analogous to the pictorial description of the Kitaev chain problem \cite{KitaevPU2001}.      

\textit{Comparison of PDI with the scattering matrix invariant and topological visibility:} As a final test, we compare the PDI with the scattering invariant \cite{Fulga_2011,Fulga_2012,Akhmerov_2011,JayDSau2016} commonly used to investigate Majorana physics in one-dimensional superconductors. We show that the PDI is free of the many biases affecting the scattering matrix invariant \cite{Day2025} and thus should be a more reliable indicator to analyze the experimental phase diagram \cite{PhysRevB.107.245423}. For example, the scattering matrix invariant of a clean finite wire supporting a pair of Majorana end modes indicates a topologically trivial phase if the effective coupling to the leads is less than the energy splitting of the Majorana modes associated with their small spatial overlap \cite{Day2025}. By construction, this issue does not affect the PDI, as it represents the topological invariant of the (clean) infinite wire obtained by periodically repeating the finite system. The weak tunneling bias can be mitigated by considering dissipative terms associated with the presence of quasiparticle sources or sinks \cite{Microsoft2025,pikulin2021protocolidentifytopologicalsuperconducting}. However, in the presence of disorder the corresponding ``invariant'' --- the so-called called topological visibility (TV) --- can generate false negatives (due to weak coupling to the Majorana modes) and false positives (due to the presence of partially separated ABSs) even in long wires. As an example, consider the topological phase diagram of a disordered $16~\muup$m wire in Fig. \ref{Fig2}(b) and the corresponding TV map in Fig. \ref{Fig2}(c). To identify the source of the manifest discrepancy between the two maps, we calculate the DOS and LDOS for specific control parameter values. The low-energy DOS along the $\mu=1.1~$meV cut is shown in Fig. \ref{Fig4}(b). While the PDI predicts a trivial phase, the TV has negative values (close to $-1$) above $\Gamma\approx 0.73~$meV, suggesting a topological SC phase [see Fig. \ref{Fig4}(a)]. The finite energy splitting of the low-energy modes revealed by the DOS, which contrasts sharply with the (practically) zero-energy Majorana feature in Fig. \ref{Fig3}, already suggests that these low-energy modes do not correspond to well-separated Majoranas. This is confirmed by the LDOS calculations, e.g., the result shown in Fig. \ref{Fig4}(c) showing an ABS consisting of a pair of partially-separated Majorana modes localized near the left end of the wire. Similar results corresponding to different disorder realizations and control parameter values confirm that, sufficiently far from a phase boundary, a nontrivial value of the PDI is generically associated with the presence of a pair of well-separated Majorana modes, hence consistent with the possibility of defining operationally a topological phase. By contrast, the scattering invariant is, in general, an unreliable topological indicator.      

\textit{Summary and Conclusion:} We propose a general approach to defining reliable topological indicators for finite-size systems with disorder by constructing the topological invariant of the infinite system obtained by periodically repeating the original disordered system. Using the one-dimensional SM-SC based topological superconductor as an example, we explicitly construct the proposed {\em periodic disorder invariant} (PDI) and demonstrate that it provides a reliable topological indicator free from the biases affecting other indicators, such as, e.g., the scattering matrix invariant. 

The concatenation procedure underlying the PDI has a transparent physical interpretation: the system is topologically nontrivial if it supports near-zero-energy boundary modes with the property that their inter-cell hybridization is stronger than their intra-cell coupling.  This ensures that the low-energy modes at the boundaries of the periodically repeated finite cells are gapped out, leaving only the low-energy modes at the two ends of the repeated system. This depiction of the topological phase closely resembles the pictorial representation of the topological phase of the 1D Kitaev lattice\cite{KitaevPU2001}. The general concept of the PDI can be extended to higher dimensions and non-superconducting topological systems as well. For example, for a disordered 2D topological system with chiral edge states, constructing a superlattice out of a cluster of plaquettes results, in the topological phase, in all inter-plaquette boundaries becoming gapped, while the global (outer) boundary remaining gapless. By contrast, a trivial PDI value indicates that the cluster supports weakly coupled low-energy {\em local} loop modes. 

By circumventing the intrinsic ambiguities associated with finite size topological indicators, the proposed PDI approach is capable of providing valuable insights into the topological properties of disordered systems, particularly in the strong disorder limit. In the case of the 1D topological superconductor, for example, the proposed approach reveals the ``shift'' of the topological region towards larger values of chemical potential and Zeeman field (Fig. 1), the relationship between the ``fragmentation'' of the topological phase diagram and finite size effects (Fig. 2), and the meaning of the robustness of the (operationally-defined) topological phase against increasing the system size. In particular, in view of the comparison with the scattering matrix invariant and the topological visibility indicator, we show that the PDE topological indicator proposed in this paper is free from the biases affecting other commonly used finite-size indicators, and captures the nature (topological or trivial) of the phase at generic points in parameter space, providing a more reliable tool for interpreting experimental results such as, for example, those in Ref.~\onlinecite{PhysRevB.107.245423}.

\textit{Acknowledgment:}
T.D.S. and B.E. thank ONR-N000142312061 for support. S.T. and B.B.R. acknowledge support from SC Quantum, ARO W911NF2210247, and ONR-N000142312061. S.T. thanks Jay D. Sau for fruitful discussions. B.B.R. thanks M. Wimmer for fruitful discussions. 

\bibliography{topoinvariant_references}

\end{document}